\documentclass[preprint,preprintnumbers,amsmath,amssymb]{revtex4}

\usepackage{graphicx}
\usepackage{rotating}
\usepackage{dcolumn}
\usepackage{bm}

\begin{document}


\title{A coupled Volterra system and its exact solutions}
\author{S. Y. Lou$^{1,2}$, Bin Tong$^1$, Man Jia$^{2}$ and Jin-hua Li$^{2}$}
\affiliation{\small $^{1}$Department of Physics, Shanghai Jiao
Tong University, Shanghai, 200030, China\\ \small $^{2}$Department
of Physics, Ningbo University, Ningbo, 315211, China}

\begin{abstract}
A coupled Volterra system is proposed. The model can be considered
as one of the integrable discrete form of the coupled integrable
KdV system which is a significant physical model. Many types of
cnoidal waves, positons, negatons (solitons) and complexitons of
the model are obtained by a simple rational expansion method of
the Jacobi elliptic functions, trigonometric functions and
hyperbolic functions.
\end{abstract}


\maketitle

\section{Introduction.}

The Volterra system \cite{volt}
\begin{eqnarray}
a_{nt}-a_n(a_{n-1}-a_{n+1})=0, \label{volt}
\end{eqnarray}
is one of the famous integrable differential-difference systems
which has been applied in various physical systems such as the
network, statistical physics and biology \cite{volt1}. Some types
of exact solutions of the model have been studied by many authors
(say \cite{Zhu}).

It is also interesting that the Volterra system is one of the
simplest discrete form of the KdV equation. In fact, if we write
\begin{eqnarray}
a_n&=&1+\delta^2u\left((n-2t)\delta,\ \frac13\delta^3t\right)+O(\delta^3)\\
&\equiv& 1+\delta^2u(x,\ \tau)+O(\delta^3),\\
a_{n\pm1}&=&1+\delta^2u(x\pm \delta,\ \tau)+O(\delta^3),
\end{eqnarray}
then \eqref{volt} becomes the well known KdV equation
\begin{eqnarray}
\frac13 (u_\tau +6uu_x+u_{xxx})\delta^5+O(\delta^6)=0.\label{KdV}
\end{eqnarray}

Recently, some types of integrable coupled KdV system have been
derived from some different physical fields including the
atmospheric dynamics \cite{cKdV}, Bose-Einstein condensation
\cite{BEC} and two-wave modes in a shallow stratified liquid
\cite{cKdV1}. A common special interesting case of
\cite{cKdV}--\cite{cKdV1} has the following form
\begin{eqnarray}
&&u_t+6\alpha vv_x+6uu_x+u_{xxx}=0,\label{ckdv1}\\
&&v_t+6vu_x+6uv_x+v_{xxx}=0.\label{ckdv2}
\end{eqnarray}
Some kinds of analytical negatons, positons and complexitons of
the coupled KdV system \eqref{ckdv1}--\eqref{ckdv2} for
$\alpha=-1$ are studied in \cite{Hu}.

On the other hand, with the development of the computer science
and the discreteness of the micro physics, to look for the
integrable discrete forms of the useful continuous integrable
models becomes a hot topic in nonlinear science. Actually, one
continuous integrable model may have some different integrable
discrete forms. For instance, the integrable discrete versions of
the KdV equation may be the Volterra equation, the Toda lattice
\cite{toda}, the Ablowitz model \cite{ab} and the hybrid lattice
\cite{hyb}, etc.

In section II, we propose a coupled Volterra system which is a
discrete version of the coupled KdV system
\eqref{ckdv1}--\eqref{ckdv2} and study the integrability of the
coupled Volterra system. The periodic cnoidal waves, solitons
(negatons), positons and complexitons are investigated by using
some suitable solution ansatzs in section III and section IV for
$\alpha>0$ and $\alpha<0$, respectively. A short summary is
presented in the last section.

\section{A coupled Volterra system}

It is interesting that the following coupled Volterra system
\begin{eqnarray}
&&a_{nt}-a_n(a_{n-1}-a_{n+1})-\alpha b_n(b_{n-1}-b_{n+1})=0,\label{cvolt1}\\
&&b_{nt}-a_n(b_{n-1}-b_{n+1})-b_n(a_{n-1}-a_{n+1})=0,
\label{cvolt2}
\end{eqnarray}
is an integrable extension of the usual Volterra system
\eqref{volt}. It is obvious that both $b_n=0$ and
$b_n=\frac1{\sqrt{\alpha}}a(n,t)$ reduce the coupled Volterra
system \eqref{cvolt1}--\eqref{cvolt2} to the usual Volterra
equation \eqref{volt}.

The integrability of the coupled Volterra system
\eqref{cvolt1}--\eqref{cvolt2} is guaranteed by the following
theorem. \\
{\bf Theorem. \rm \em The coupled Volterra system
\eqref{cvolt1}--\eqref{cvolt2} possesses the following Lax pair,
\begin{eqnarray}
&&L\psi_n=\Lambda\psi_n,\label{lax1}\\
&&\psi_{nt}=M\psi_n, \label{lax2}
\end{eqnarray}
where
\begin{eqnarray}
L&\equiv&\left(\begin{array}{cc}a_nT_++T_- & \alpha b_n T_+\\
b_n T_+& a_nT_++T_-
  \end{array}\right),\quad \Lambda=\left(\begin{array}{cc}\lambda_1 & \alpha \lambda_2\\
\lambda_2& \lambda_1
  \end{array}\right),\quad \psi_n\equiv \left(\begin{array}{c}\psi_{1n} \\
\psi_{2n}
  \end{array}\right),\label{L}\\
M&\equiv&-\left(\begin{array}{cc}a_na_{n+1}+\alpha b_nb_{n+1} & \alpha (a_nb_{n+1}+ b_na_{n+1})\\
a_nb_{n+1}+ b_na_{n+1} & a_na_{n+1}+\alpha
b_nb_{n+1}\end{array}\right)T_+^2, \label{M}
\end{eqnarray}
and  $T_+$ and $T_-$ are shift operators defined by
\begin{eqnarray}
T_+f_n\equiv f_{n+1},\ T_-f_n\equiv f_{n-1} \label{T}
\end{eqnarray}
for an arbitrary function $f_n$.}\\
\bf Proof. \rm By the direct calculations, one can prove that the
matrices $M$ and $\Lambda$ are commutable and then the
compatibility condition of \eqref{L} and  \eqref{M}  is just the
zero curvature condition
\begin{eqnarray}
L_t+LM-ML=0. \label{zero}
\end{eqnarray}
Substituting the definition equations of $L$ and $M$ into
\eqref{zero} just leads to the coupled Volterra system
\eqref{cvolt1}--\eqref{cvolt2}. The theorem is proved.

Another interesting fact is that the coupled Volterra system
\eqref{cvolt1}--\eqref{cvolt2} is really a discrete form of the
coupled KdV system \eqref{ckdv1}--\eqref{ckdv2}. Applying
 the following continuous limiting
procedure
\begin{eqnarray}
a_n&=&1+\delta^2u\left((n-2t)\delta,\ \frac13\delta^3t\right)+O(\delta^3)\nonumber\\
&\equiv& 1+\delta^2u(x,\ \tau)+O(\delta^3),\label{an}\\
b_n&=&\delta^2 v\left((n-2t)\delta,\
\frac13\delta^3t\right)+O(\delta^3)\nonumber\\
&\equiv&\delta^2 v\left(x,\ \tau\right)+O(\delta^3),\\
a_{n\pm1}&=&1+\delta^2u(x\pm \delta,\ \tau)+O(\delta^3),\\
b_{n\pm1}&=&\delta^2v(x\pm \delta,\ \tau)+O(\delta^3)\label{bn}
\end{eqnarray}
to \eqref{cvolt1}--\eqref{cvolt2}, we have
\begin{eqnarray}
&&\frac13(u_\tau+6\alpha vv_x+6uu_x+u_{xxx})\delta^5+O(\delta^6)=0,\label{Ckdv1}\\
&&\frac13(v_\tau+6vu_x+6uv_x+v_{xxx})\delta^5+O(\delta^6)=0\label{Ckdv2}
\end{eqnarray}
which is just the special coupled KdV system
\eqref{ckdv1}--\eqref{ckdv2} with $t\rightarrow \tau$.

\section{cnoidal waves, solitons and positons of the coupled Volterra
system for $\alpha>0$}

In the continuous case, the rational expansion of the elliptic and
hyperbolic functions is one of the simplest methods to find
travelling periodic and solitary wave solutions. Fortunately, this
method is valid also in discrete cases.

To find some types of cnoidal wave solutions of the coupled
Volterra system \eqref{cvolt1}--\eqref{cvolt2}, one may take some
types of the elliptic function expansion ansatzs, say,
\begin{eqnarray}
a(n,t)&=&\frac{P(f,g,h)}{R(f,g,h)},\label{ansatz1}\\
b(n,t)&=&\frac{Q(f,g,h)}{R(f,g,h)}\label{ansatz2}
\end{eqnarray}
where $\{P(f,g,h),\ Q(f,g,h), R(f,g,h)\}$ are polynomial functions
of $\{f,\ g,\ h\}$ and $f\equiv \mbox{\rm sn}(kn+ct,m)$, $g\equiv
\mbox{\rm cn}(kn+ct,m)$ and $h\equiv \mbox{\rm dn}(kn+ct,m)$ are
Jacobi elliptic functions with three constant parameters, the wave
number $k$, the angular frequency $\omega$ and the modulus $m$.
Because of the computational difficulty, here, we just give some
special examples of \eqref{ansatz1}--\eqref{ansatz2}.\\
\bf Case 1. \rm The first simple expansion ansatz reads
\begin{eqnarray}
a(n,t)&=&\frac{a_1+c_1 \mbox{\rm cn}^2(kn+ct,\ m)}{a_2+c_2 \mbox{\rm cn}^2(kn+ct,\ m)},\label{ep1}\\
b(n,t)&=&\frac{A[1+a_0 \mbox{\rm cn}^2(kn+ct,\ m)]}{a_2+c_2
\mbox{\rm cn}^2(kn+ct,\ m)}\label{ep2}
\end{eqnarray}
where $a_1,\ c_1,\ a_2,\ c_2,\ A,\ k,\ m$ and $c$ are constants
that should be determined later.

Substituting \eqref{ep1}--\eqref{ep2} into
\eqref{cvolt1}--\eqref{cvolt2} and vanishing the coefficients of
the different powers of the Jacobi elliptic function $\mbox{\rm
cn}^2(kn+ct)$, we can obtain a complicated determining equation
system for the undetermined constants. Fortunately, there exist a
unique general solution for the determining equation system. The
result reads ($s\equiv \mbox{\rm sn}(k\delta,\ m),\ d\equiv
\mbox{\rm dn}(k\delta,\ m),\ C\equiv \mbox{\rm cn}(k\delta,\ m)$)
\begin{eqnarray}
s^2&=&\frac{4a_2c_2(a_2+c_2)[m^2(a_2+c_2)-c_2]}{[m^2(a_2+c_2)^2-c_2^2]^2},\label{s2}\\
A^2&=&\frac{a_2c_2c^2(a_2+c_2)c^2[m^2(a_2+c_2)-c_2]}{\alpha (a_2a_0-c_2)^2},\label{a2}\\
a_1&=&\frac{cm^2s(a_2+c_2)(a_2-c_2+2a_0a_2)[(a_2+c_2)m^2s^2-2c_2]}{4dCc_2(a_0a_2-c_2)}-\frac{ca_2}{2dCs}\nonumber\\
&&
+\frac{cc_2s(2a_0a_2-c_2)(2a_2+2c_2-c_2s^2)}{4dC(a_2+c_2)(a_0a_2-c_2)},\\
c_1&=&\frac{a_1(a_0c_2+2c_2-a_0a_2)}{a_2+2a_0a_2-c_2}+\frac{c(a_2+c_2-c_2s^2)^2(c_2-a_0a_2)}
{2dCs(a_2+c_2)(a_2+2a_0a_2-c_2)},
\end{eqnarray}
while the constants $a_2,\ c_2,\ c,\ a_0$ and $m$ remain to be
free parameters.

From \eqref{s2} and \eqref{a2}, it is known that the real
condition of the periodic solution requires
$$\alpha>0.$$

Fig. 1 shows the structure of the periodic solution \eqref{ep1}
with the parameter selections
\begin{eqnarray}
\alpha=c=a_2=c_2=1,\ a_0=2,\ \delta=1, m=0.999 \label{para1}
\end{eqnarray}
at time $t=0$.

\input epsf
\begin{figure}
\epsfxsize=7cm\epsfysize=5cm\epsfbox{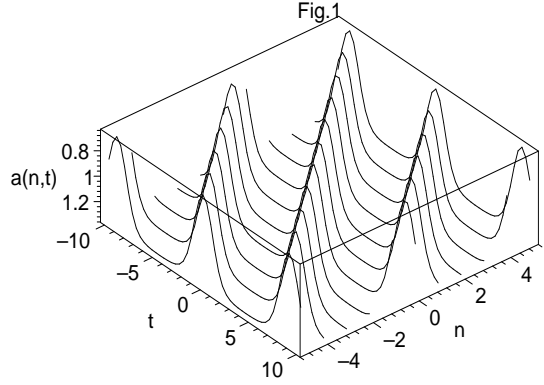}
 \caption{The structure of the periodic wave expressed by \eqref{ep1} with the parameters \eqref{para1} at time $t=0$.}
\end{figure}

It is remarkable that for the $\alpha>0$ case, in addition to the
above cnoidal wave solution, we can find many other nonequivalent
periodic waves by similar procedures. In the following, we just
list some of them. \\
\bf Case 2. \rm
\begin{eqnarray}
&&a(n,t)=\frac{a_1+c_1 \mbox{\rm cn}(kn+ct,\ m)}{a_2+c_2 \mbox{\rm cn}(kn+ct,\ m)},\label{2.1}\\
&&b(n,t)=\frac{A_1+A_2 \mbox{\rm cn}(kn+ct,\ m)]}{a_2+c_2
\mbox{\rm cn}(kn+ct,\ m)}\label{2.2}
\end{eqnarray}
where the constants $a_1,\ c_1,\ a_2,\ c_2,\ A_1,\ A_2,\ k,\ c$
and $m$ satisfy the following conditions:
\begin{eqnarray}
&&a_1=\frac{sc}{4a_2d}(m^2a_2^4+1-m^2),\ \label{2.3}\\
&&\frac{A_2}{A_1}=\frac{2a_2\big[c+2c_1sd-cm^2s^2(1-a_2^2)\big]}{cs^2\big[m^2(1-a_2^2)^2-1\big]+2a_2^2(c+2c_1sd)},\\
&&A_1=\pm \frac{cs^2\big[m^2(1-a_2^2)^2-1\big]+2a_2^2(c+2c_1sd)}{4sda_2\sqrt{\alpha}}\\
&&C=1+\frac{s^2}{2a_2^2}\left[m^2(1-a_2^2)^2-1\right],\ c_2=1.
\label{2.4}
\end{eqnarray}
\bf Case 3. \rm
\begin{eqnarray}
&&a(n,t)=\frac{a_1+c_1 \mbox{\rm dn}(kn+ct,\ m)}{a_2+c_2 \mbox{\rm dn}(kn+ct,\ m)},\label{3.1}\\
&&b(n,t)=\frac{A_1+A_2 \mbox{\rm dn}(kn+ct,\ m)]}{a_2+c_2
\mbox{\rm dn}(kn+ct,\ m)}\label{3.2}
\end{eqnarray}
with the constant constraints
\begin{eqnarray}
&&a_1=\frac{sc}{4a_2C}(m^2+a_2^4-1)+c_1a_2,\ \label{3.3}\\
&&\frac{A_2}{A_1}=\frac{2a_2\big[c+2c_1sC-cs^2(1-a_2^2)\big]}{cs^2\big[(1-a_2^2)^2-m^2\big]+2a_2^2(c+2c_1sC)},\\
&&A_1=\pm \frac{cs^2\big[(1-a_2^2)^2-m^2\big]+2a_2^2(c+2c_1sC)}{4sCa_2\sqrt{\alpha}}\\
&&d=1+\frac{s^2}{2a_2^2}\left[(1-a_2^2)^2-m^2\right],\ c_2=1.
\label{3.4}
\end{eqnarray}
\bf Case 4. \rm
\begin{eqnarray}
&&a(n,t)=\frac{a_1+c_1 \mbox{\rm sn}(kn+ct,\ m)}{a_2+c_2 \mbox{\rm sn}(kn+ct,\ m)},\label{4.1}\\
&&b(n,t)=\frac{A_1+A_2 \mbox{\rm sn}(kn+ct,\ m)]}{a_2+c_2
\mbox{\rm sn}(kn+ct,\ m)}\label{4.2}
\end{eqnarray}
with the constant constraints
\begin{eqnarray}
&&c_1=\frac{a_1}{a_2}-\frac{c(Cd-1)}{2s(1+m^2a_2^4)}(m^2a_2^4-1),\ \label{4.3}\\
&&\frac{A_2}{A_1}=\frac{(1+m^2a_2^4)(2sa_1+cdCa_2)}{a_2\big[m^2a_2^4(ca_2+2sa_1)+2sa_1+2cdCa_2-ca_2\big]},\\
&&A_1=\pm \frac{m^2a_2^4(ca_2+2sa_1)+2sa_1+2cdCa_2-ca_2}{2a_2\sqrt{2\alpha(1-dC)(1+a_2^4m^2)}}\\
&&s^2=\frac{2a_2^2(1-d C)}{1+m^2a_2^4},\ c_2=1. \label{4.4}
\end{eqnarray}
\bf Case 5. \rm
\begin{eqnarray}
&&a(n,t)=\frac{a_1\mbox{\rm cn}(kn+ct,\ m)+c_1 \mbox{\rm sn}(kn+ct,\ m)}
{a_2\mbox{\rm cn}(kn+ct,\ m)+c_2 \mbox{\rm sn}(kn+ct,\ m)},\label{5.1}\\
&&b(n,t)=\frac{A_1\mbox{\rm cn}(kn+ct,\ m)+A_2 \mbox{\rm
sn}(kn+ct,\ m)}{a_2\mbox{\rm cn}(kn+ct,\ m)+c_2 \mbox{\rm
sn}(kn+ct,\ m)}\label{5.2}
\end{eqnarray}
with the constant constraints
\begin{eqnarray}
&&c_1=\frac{a_1}{a_2}-\frac{c(d-1)(m^2a_2^4-a_2^4+1)}{2sC[m^2a_2^4-(1+a_2^2)^2]},\ \label{5.3}\\
&&\frac{A_2}{A_1}=\frac{[(1+a_2^2)^2-m^2a_2^4](cda_2+2sCa_1)}{a_2\big[m^2a_2^4(ca_2+2sa_1C)
-(1+a_2^2)(2sCa_1+2cda_2-ca_2+ca_2^3+2sCa_1a_2^2)\big]},\\
&&A_1=\pm \frac{m^2a_2^4(ca_2+2sa_1C)-(1+a_2^2)(2sCa_1+2cda_2-ca_2+ca_2^3+2sCa_1a_2^2)}{4a_2C(1-d)\sqrt{\alpha}}\\
&&s^2=\frac{2a_2^2(1-d)}{(1+a_2^2)^2-m^2a_2^4},\ c_2=1.
\label{5.4}
\end{eqnarray}
\bf Case 6. \rm
\begin{eqnarray}
&&a(n,t)=\frac{a_1\mbox{\rm dn}(kn+ct,\ m)+c_1 \mbox{\rm
sn}(kn+ct,\ m)}
{a_2\mbox{\rm dn}(kn+ct,\ m)+c_2 \mbox{\rm sn}(kn+ct,\ m)},\label{6.1}\\
&&b(n,t)=\frac{A_1\mbox{\rm dn}(kn+ct,\ m)+A_2 \mbox{\rm
sn}(kn+ct,\ m)}{a_2\mbox{\rm dn}(kn+ct,\ m)+c_2 \mbox{\rm
sn}(kn+ct,\ m)}\label{6.2}
\end{eqnarray}
with the constant constraints
\begin{eqnarray}
&&c_1=\frac{a_1}{a_2}-\frac{c(C-1)(m^4a_2^4-1-m^2a_2^4)}{2sd[(1+m^2a_2^2)^2-m^2a_2^4]},\ \label{6.3}\\
&&\frac{A_2}{A_1}=\frac{[(1+m^2a_2^2)^2-m^2a_2^4](cCa_2+2sda_1)}{a_2\big[2sda_1[(1+m^2a_2^2)^2-m^2a_2^4]
+ca_2[m^4a_2^4+m^2a_2^2(2C-a_2^2)+2C-1]\big]},\\
&&A_1=\pm \frac{s\big[2sda_1[(1+m^2a_2^2)^2-m^2a_2^4]+ca_2[m^4a_2^4+m^2a_2^2(2C-a_2^2)
+2C-1]\big]}{4a_2^2d(1-C)\sqrt{\alpha}},\\
&&s^2=\frac{2a_2^2(1-C)}{(1+a_2^2m^2)^2-m^2a_2^4},\ c_2=1.
\label{6.4}
\end{eqnarray}
\bf Case 7. \rm
\begin{eqnarray}
&&a(n,t)=\frac{a_1\mbox{\rm dn}(kn+ct,\ m)+c_1 \mbox{\rm
cn}(kn+ct,\ m)}
{a_2\mbox{\rm dn}(kn+ct,\ m)+c_2 \mbox{\rm cn}(kn+ct,\ m)},\label{7.1}\\
&&b(n,t)=\frac{A_1\mbox{\rm dn}(kn+ct,\ m)+A_2 \mbox{\rm
cn}(kn+ct,\ m)}{a_2\mbox{\rm dn}(kn+ct,\ m)+c_2 \mbox{\rm
cn}(kn+ct,\ m)}\label{7.2}
\end{eqnarray}
with the constant constraints
\begin{eqnarray}
&&c_1=\frac{a_1}{a_2}-\frac{c(Cd-1)(m^2a_2^4-1)}{2s(1+m^2a_2^4)},\ \label{7.3}\\
&&\frac{A_2}{A_1}=\frac{(1+m^2a_2^4)(cCda_2+2sa_1)}{a_2\big[m^2a_2^4(ca_2+2sa_1)+2sa_1+2cda_2C-ca_2\big]},\\
&&A_1=\pm
\frac{m^2a_2^4(ca_2+2sa_1)+2sa_1+2cda_2C-ca_2}{2a_2\sqrt{2\alpha(1-d C)(1+m^2a_2^4)}},\\
&&s^2=\frac{2a_2^2(1-d C)}{1+a_2^4m^2},\ c_2=1. \label{7.4}
\end{eqnarray}

Especially, when $m\rightarrow 1$, the previous periodic wave
solutions become negaton (soliton) solutions. For instance,
\eqref{ep1} and \eqref{ep2} become the soliton solution
\begin{eqnarray}
a(n,t)&=&\frac{a_1+c_1 \mbox{\rm sech}^2(kn+ct)}{a_2+c_2 \mbox{\rm sech}^2(kn+ct)},\label{sol1}\\
b(n,t)&=&\frac{A[1+a_0 \mbox{\rm sech}^2(kn+ct)]}{a_2+c_2
\mbox{\rm sech}^2(kn+ct)}\label{sol2}
\end{eqnarray}
with
\begin{eqnarray}
S^2&=&\frac{4c_2(a_2+c_2)}{(a_2+2c_2)^2},\quad A_1^2=\frac{c_2a_2^2c^2(a_2+c_2)}{\alpha (a_2a_0-c_2)^2},\\
a_1&=&\frac{2ca_2c_2(a_2+c_2)}{S(a_2a_0-c_2)(a_2+2c_2)}-\frac{a_2^3c}{2S(a_2a_0-c_2)^2},\\
c_1&=&\frac{2cc_2^2(a_2+c_2)}{S(a_2a_0-c_2)(a_2+2c_2)}-\frac{c_2c(5a_2^2+12a_2c_2+8c_2^2)}{2S(a_2a_0-c_2)^2},
\end{eqnarray}

Fig. 2 shows the structure of the soliton solution \eqref{ep1}
with the same parameter selections as \eqref{para1} except for
$m=1$.
\input epsf
\begin{figure}
\epsfxsize=7cm\epsfysize=5cm\epsfbox{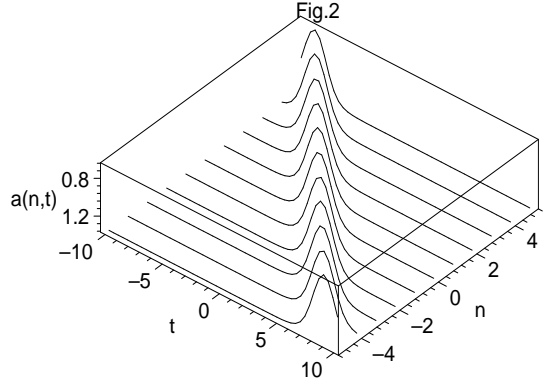}
 \caption{The structure of the solitary wave expressed by \eqref{sol1} which is the limit case of the figure 1 for
 $m=1$.}
\end{figure}

In \cite{Hu}, for the $\alpha<0$ case, it is found that the
coupled KdV system \eqref{ckdv1} and \eqref{ckdv2} possesses not
only the analytic solitons (negatons) but also analytical positons
and complexitons. Now the natural question is whether the coupled
Volterra system possesses analytical positons and/or complexitons.

Similarly, all the cnoidal wave solutions presented above will
reduced to the positon solutions when we take $m=0$. For instance,
the first type of the positon solutions can be obtained from the
periodic solution \eqref{ep1} and \eqref{ep2} by taking $m=0$,
which has the form of
\begin{eqnarray}
a(n,\ t)&=&\frac{2[a_1+c_1\cos^2(kn+ct)]}{\cos(2kn+2ct)\pm \cos(k\delta)},\label{posi1a}\\
b(n,\
t)&=&\frac{2A_1[1+a_0\cos^2(kn+ct)]}{\sqrt{\alpha}[\cos(2kn+2ct)\pm
\cos(k\delta)]},\label{posi1b}
\end{eqnarray}
where
\begin{eqnarray}
a_1&=&\frac{c\left[4+a_0\big(2+\cos(2k\delta)+\cos(4k\delta)\big)\mp
(1+a_0)\cos(k\delta)\cos(2k\delta)\right]}{4\sin(2k\delta)[2+a_0\mp
a_0\cos(k\delta)]},\\
c_1&=&-\frac{c[2+a_0\mp
a_0\cos(k\delta)\cos(2k\delta)]}{\sin(2k\delta)[2+a_0\mp
a_0\cos(k\delta)]},\\
A_1^2&=&\frac{\sin(k\delta)^2}{[2+a_0\mp a_0\cos(k\delta)]^2},
\label{posiA1}
\end{eqnarray}
and $c,\ a_0$ and $k$ are arbitrary constants.

Obviously, the positon solution \eqref{posi1a}--\eqref{posi1b} is
always singular. Fig. 3 shows a special structure of
\eqref{posi1a} and \eqref{posi1b} with the parameter selections
\begin{eqnarray}
c=\alpha=\delta=1,\ a_0=2,\ k=0.6435011088.\label{paraposi}
\end{eqnarray}
Actually, all the positon solutions obtained from Cases 1 to 7 by
taking $m=0$ are singular except for the trivial constant solution
in Case 3.

\input epsf
\begin{figure}
\epsfxsize=7cm\epsfysize=5cm\epsfbox{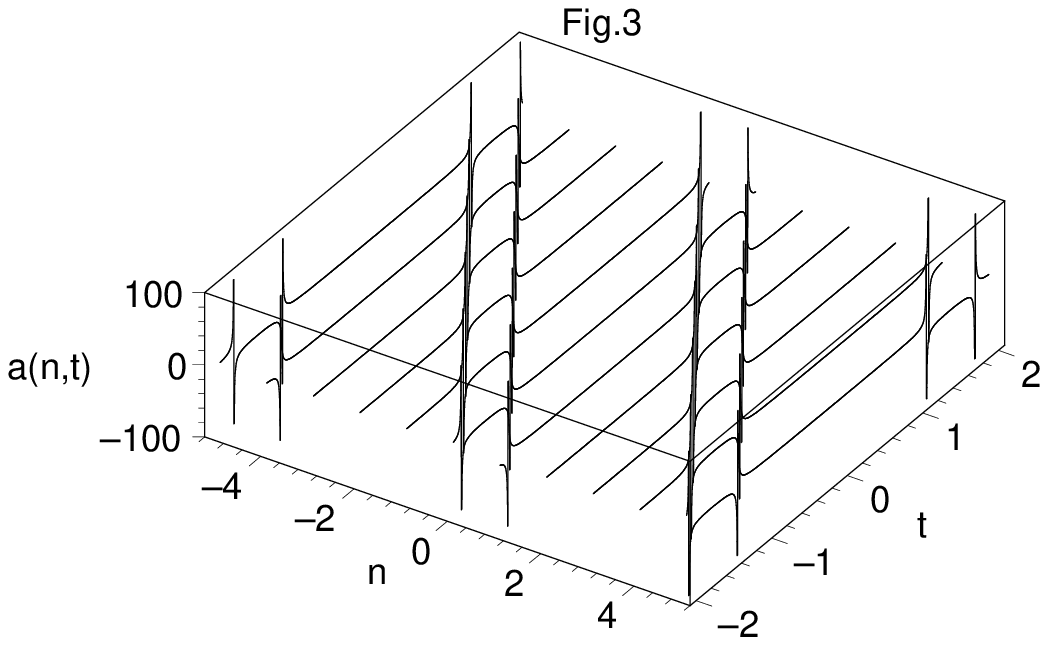}
 \caption{A typical positon structure expressed by \eqref{posi1a}--\eqref{posiA1}
 with the parameter selections \eqref{paraposi}.}
\end{figure}

For a coupled nonlinear system, there may be different types of
solitons and positons. The first type of soliton solutions given
by \eqref{sol1} and \eqref{sol2} possesses the property that the
 fields $a(n,\ t)$ and $b(n,\ t)$ both have the ring or bell shape.
The coupled Volterra system can have other kinds of soliton
solutions. For instance, by substituting the following solution
ansatz
\begin{eqnarray}
a(n,t)&=&\frac{a_0+a_1\cosh(kn+ct)+a_2\cosh(2kn+2ct)}{b_0+b_1\cosh(kn+ct)+b_2\cosh(2kn+2ct)},\label{Sol1}\\
b(n,t)&=&\frac{d_0\sinh(kn+ct)}{b_0+b_1\cosh(kn+ct)+b_2\cosh(2kn+2ct)}\label{Sol2}
\end{eqnarray}
into the coupled Volterra system \eqref{cvolt1} and
\eqref{cvolt2}, one can find that
\begin{eqnarray}
d_0&=&\mp c\sin(c_0)\left[\cos\left(\frac32k\delta\right)-\cos\left(\frac12k\delta\right)\right],\label{D0}\\
b_0&=&\sqrt{\alpha}\sin(k\delta){(1+\cos(2c_0)+\cos(k\delta))},\\
b_1&=&\pm4\sqrt{\alpha}\sin(k\delta)\cos(c_0)\cos\left(\frac12k\delta\right),\\
a_0&=&-\frac{c}2
\left[\cos(2c_0)+2\cos\left(\frac12k\delta\right)\cos\left(\frac32k\delta\right)\right],\\
a_1&=&\mp c
\cos(c_0)\left[\cos\left(\frac12k\delta\right)+\cos\left(\frac32k\delta\right)\right],\\
b_2&=&\sqrt{\alpha}\sin(k\delta),\ a_2=-\frac c2,\label{B2}
\end{eqnarray}
where $c,\ c_0$ and $k$ are arbitrary constants.

From \eqref{D0}--\eqref{B2}, we know that this kind of soliton
solution is also valid only for $\alpha>0$. The solution
\eqref{Sol1}--\eqref{Sol2} may be singular or analytical based on
the different selections of the parameters.

Fig. 4 exhibits the structure of the soliton solution given by
\eqref{Sol1}--\eqref{B2} with the special parameter selections
\begin{eqnarray}
\delta=c_0=\alpha=1,\ c=k=2, \label{paras2}
\end{eqnarray}
for the upper sign. It is clear that in this case the soliton
structures are different for the fields $a(n,\ t)$ and $b(n,\ t)$.
The soliton structure for $a(n,\ t)$ is still a bell or ring shape
while that for $b(n,\ t)$ becomes staggered.

\input epsf
\begin{figure}
\epsfxsize=7cm\epsfysize=5cm\epsfbox{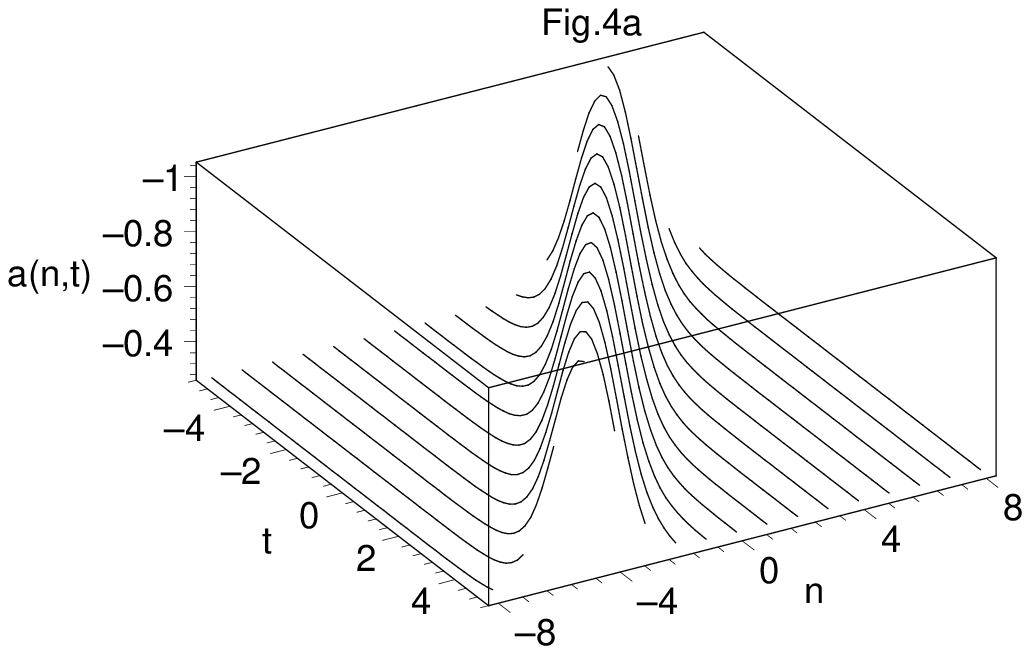}
\epsfxsize=7cm\epsfysize=5cm\epsfbox{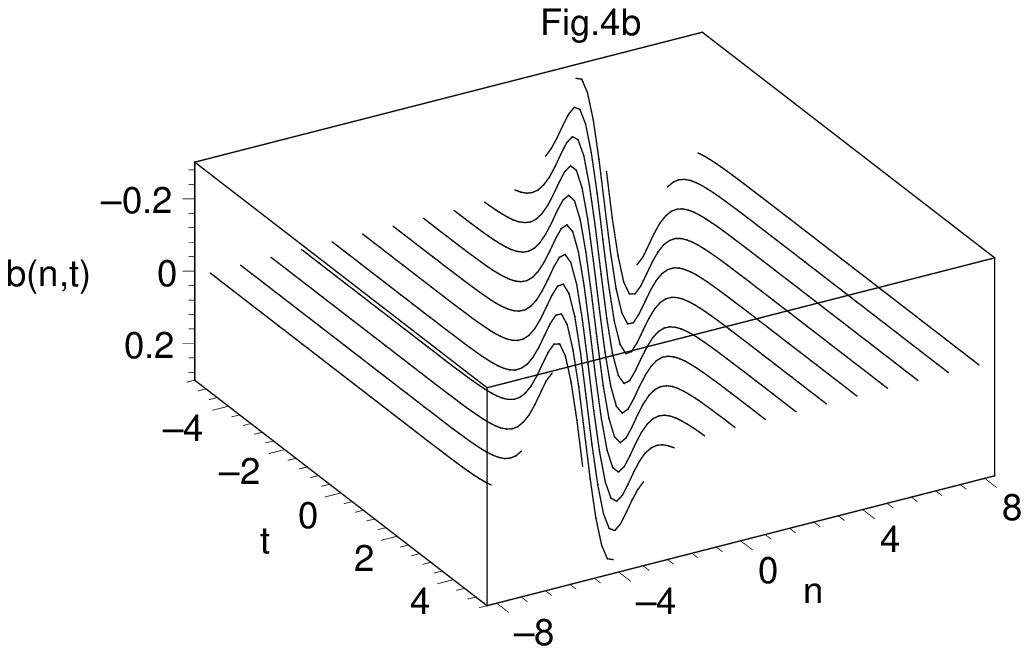}
 \caption{Second type of soliton structure expressed by \eqref{Sol1}--\eqref{B2}
 with the parameter selections \eqref{paras2}.}
\end{figure}

It is noted that because of the arbitrariness of the constants
$k,\ c$ and $c_0$, if we take
\begin{eqnarray}
k\rightarrow \sqrt{-1}k, \ c\rightarrow \sqrt{-1}c,\
c_0\rightarrow \sqrt{-1}c_0
\end{eqnarray}
then the soliton solution \eqref{Sol1}--\eqref{B2} is transformed
to another type of positon solutions
\begin{eqnarray}
a(n,t)&=&\frac{a_0+a_1\cos(kn+ct)+a_2\cos(2kn+2ct)}{b_0+b_1\cos(kn+ct)+b_2\cos(2kn+2ct)},\label{posi1}\\
b(n,t)&=&\frac{d_0\sin(kn+ct)}{b_0+b_1\cos(kn+ct)+b_2\cos(2kn+2ct)}\label{posi2}
\end{eqnarray}
with
\begin{eqnarray}
d_0&=&\mp c\sinh(c_0)\left[\cosh\left(\frac32k\delta\right)-\cosh\left(\frac12k\delta\right)\right],\label{D01}\\
b_0&=&\sqrt{\alpha}\sinh(k\delta){[1+\cosh(2c_0)+\cosh(k\delta)]},\\
b_1&=&\pm4\sqrt{\alpha}\sinh(k\delta)\cosh(c_0)\cosh\left(\frac12k\delta\right),\\
a_0&=&-\frac{c}2
\left[\cosh(2c_0)+2\cosh\left(\frac12k\delta\right)\cosh\left(\frac32k\delta\right)\right],\\
a_1&=&\mp c
\cosh(c_0)\left[\cosh\left(\frac12k\delta\right)+\cosh\left(\frac32k\delta\right)\right],\\
b_2&=&\sqrt{\alpha}\sinh(k\delta),\ a_2=-\frac c2.\label{B21}
\end{eqnarray}
Fig. 5 displays the structure of the positon solution
\eqref{posi1}--\eqref{B21} for the upper sign with the special
parameter selections
\begin{eqnarray}
\delta=\alpha=1,\ c=k=\frac{\pi}2, c_0=\frac{\pi}3.\label{paras3}
\end{eqnarray}
Evidently, in this case the positon structure is still singular.
Up to now, for the $\alpha>0$ case, we have not yet found any
analytical positons and complexitons. However, similar to the
single KdV equation and the Toda system \cite{Ma}, we believe that
the coupled Volterra system \eqref{cvolt1} and \eqref{cvolt2} for
$\alpha>0$ does not possess analytical positons and complexitons.

\input epsf
\begin{figure}
\epsfxsize=7cm\epsfysize=5cm\epsfbox{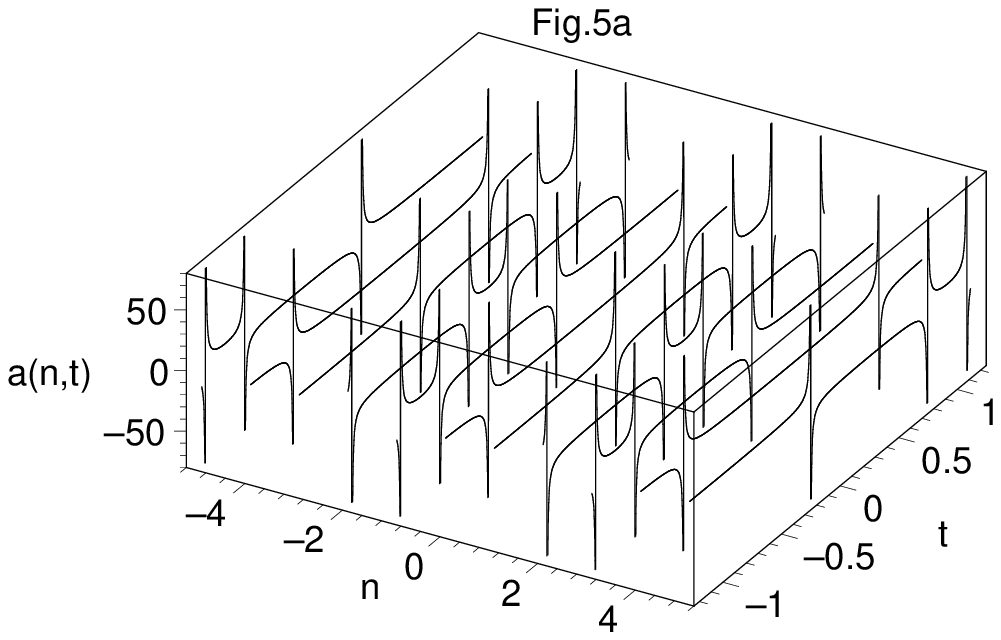}
\epsfxsize=7cm\epsfysize=5cm\epsfbox{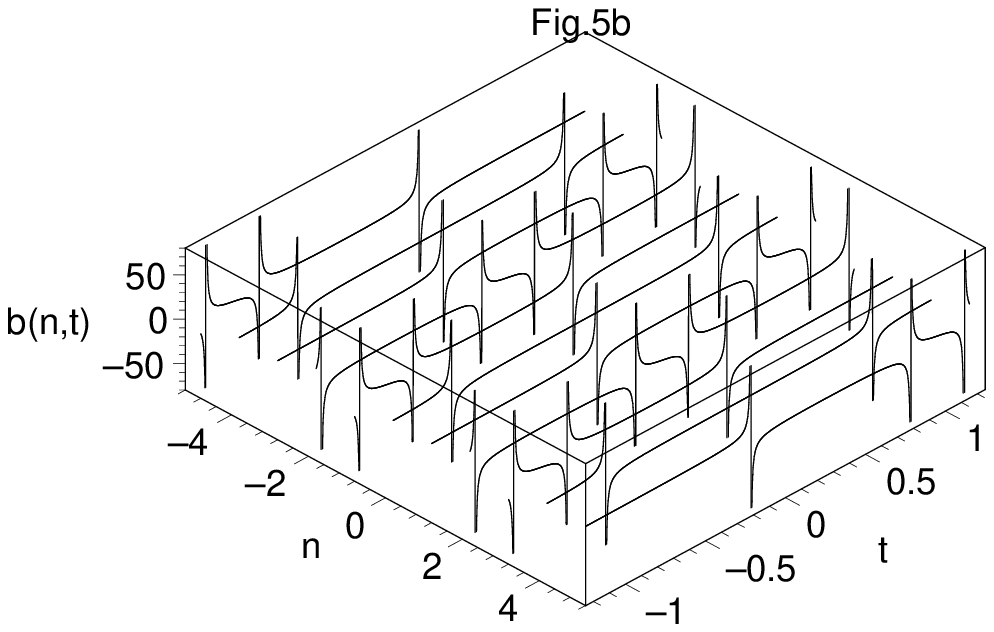}
 \caption{A typical positon structure expressed by \eqref{posi1}--\eqref{B21}
 with the parameter selections \eqref{paras3}.}
\end{figure}

\section{Solitons, positons and complexitons of the coupled Volterra
system for $\alpha<0$}

For the coupled real Volterra system \eqref{cvolt1} and
\eqref{cvolt2} with $\alpha<0$, all the cnoidal wave solution
forms in the last section are not valid, however, the function
expansion ansatz  \eqref{Sol1} and \eqref{Sol2} is still
applicable to obtain some soliton solutions.

Similar to the last section, after substituting \eqref{Sol1} and
\eqref{Sol2} into \eqref{cvolt1}--\eqref{cvolt2} for $\alpha<0$
and solving the determining equations of the parameters, one can
find that
\begin{eqnarray}
d_0&=&\frac{2c\sin(c_0)\sinh\left(\frac12k\delta\right)}{\sqrt{-\alpha}},\label{d0}\\
b_0&=&{1+\cos(2c_0)+\cosh(k\delta)},\\
b_1&=&4\cos(c_0)\cosh\left(\frac12k\delta\right),\\
a_0&=&\frac{c}{\sinh(k\delta)}
[\cos(2c_0)+\cosh(k\delta)+\cosh(2k\delta)],\\
a_1&=&\frac{c
\cos(c_0)}{\sinh(k\delta)}\left[\cosh\left(\frac12k\delta\right)+\cosh\left(\frac32k\delta\right)\right],\\
b_2&=&1,\ a_2=\frac{c}{2\sinh(k\delta)},\label{b2}
\end{eqnarray}
where $c,\ c_0$ and $k$ are arbitrary constants.

Fig. 6 shows the structure of the soliton solution expressed by
\eqref{Sol1} and \eqref{Sol2} with \eqref{d0}--\eqref{b2} and the
parameter selections
\begin{eqnarray}
c_0=\frac{\pi}2,\ c=k=\frac15,\ \delta=1,\ \alpha=-1.\label{para2}
\end{eqnarray}

\input epsf
\begin{figure}
\epsfxsize=7cm\epsfysize=5cm\epsfbox{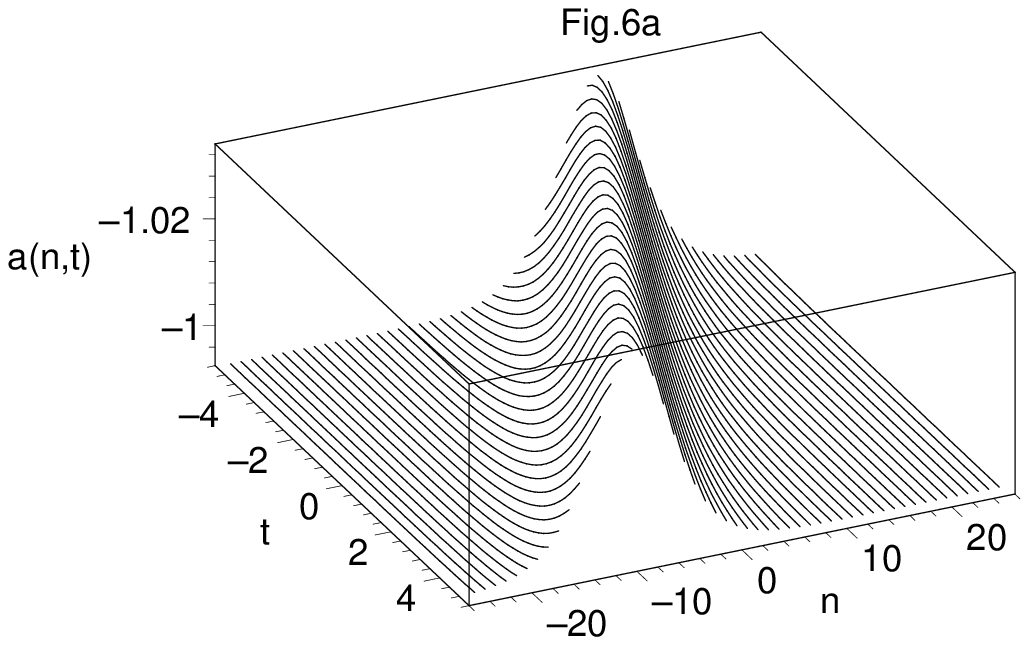}
\epsfxsize=7cm\epsfysize=5cm\epsfbox{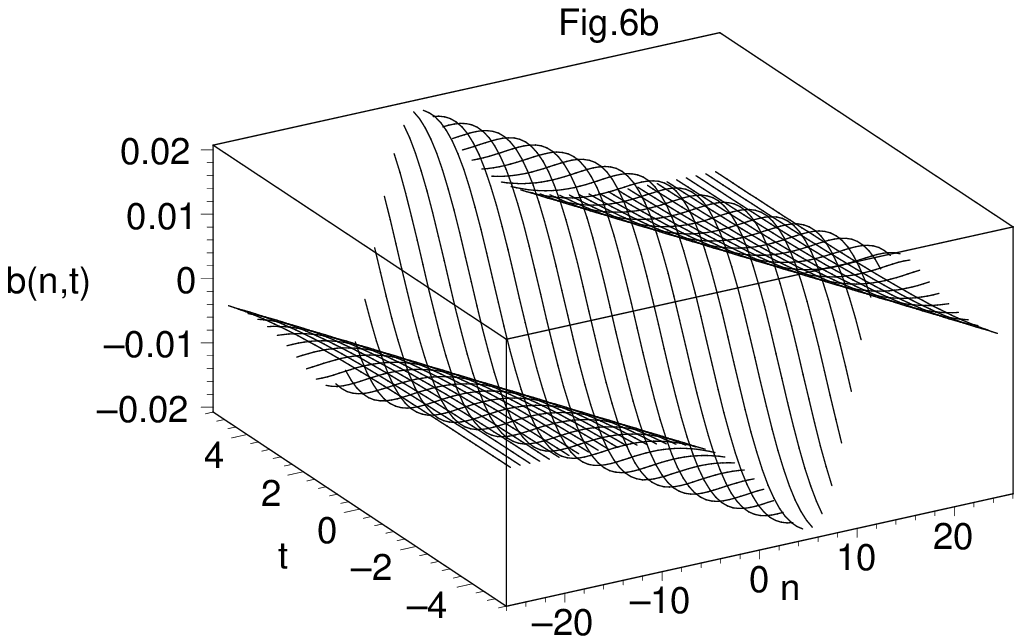}
 \caption{The structures of the soliton expressed by (a) \eqref{Sol1} and (b) \eqref{Sol2}
 with \eqref{d0}--\eqref{b2} and the parameter selections \eqref{para2}.}
\end{figure}

In the continuous case, it has been proved that the coupled KdV
system \eqref{ckdv1} and \eqref{ckdv2} with $\alpha<0$ have some
types of analytical positon and complexiton solutions \cite{Hu}.
It is interesting that in the discrete case, the coupled Volterra
system also possesses analytical positons and complexitons.

To get analytical positon solutions, one can directly apply the
constant transformations
$$k\rightarrow \sqrt{-1}k,\  c\rightarrow \sqrt{-1}c,\ c_0\rightarrow \sqrt{-1}c_0,   $$
to \eqref{Sol1}--\eqref{Sol2} with \eqref{d0}--\eqref{b2}. The
result still has the form \eqref{posi1}--\eqref{posi2} but with
the constants
\begin{eqnarray}
d_0&=&\frac{2c\sinh(c_0)\sin\left(\frac12k\delta\right)}{\sqrt{-\alpha}},\label{d01}\\
b_0&=&{1+\cosh(2c_0)+\cos(k\delta)},\\
b_1&=&4\cosh(c_0)\cos\left(\frac12k\delta\right),\\
a_0&=&-\frac{c}{2\sin(k\delta)}
[\cosh(2c_0)+\cos(k\delta)+\cos(2k\delta)],\\
a_1&=&-\frac{c
\cosh(c_0)}{\sin(k\delta)}\left[\cos\left(\frac12k\delta\right)+\cos\left(\frac32k\delta\right)\right],\\
b_2&=&1,\ a_2=-\frac{c}{2\sin(k\delta)},\label{b21}
\end{eqnarray}

Fig. 7 reveals the structure of the analytical positon solution
expressed by \eqref{Sol1} and \eqref{Sol2} with
\eqref{d01}--\eqref{b21} and the parameter selections
\begin{eqnarray}
c_0=1,\ c=1,\ k=2,\ \delta=1,\ \alpha=-1.\label{para3}
\end{eqnarray}

\input epsf
\begin{figure}
\epsfxsize=7cm\epsfysize=5cm\epsfbox{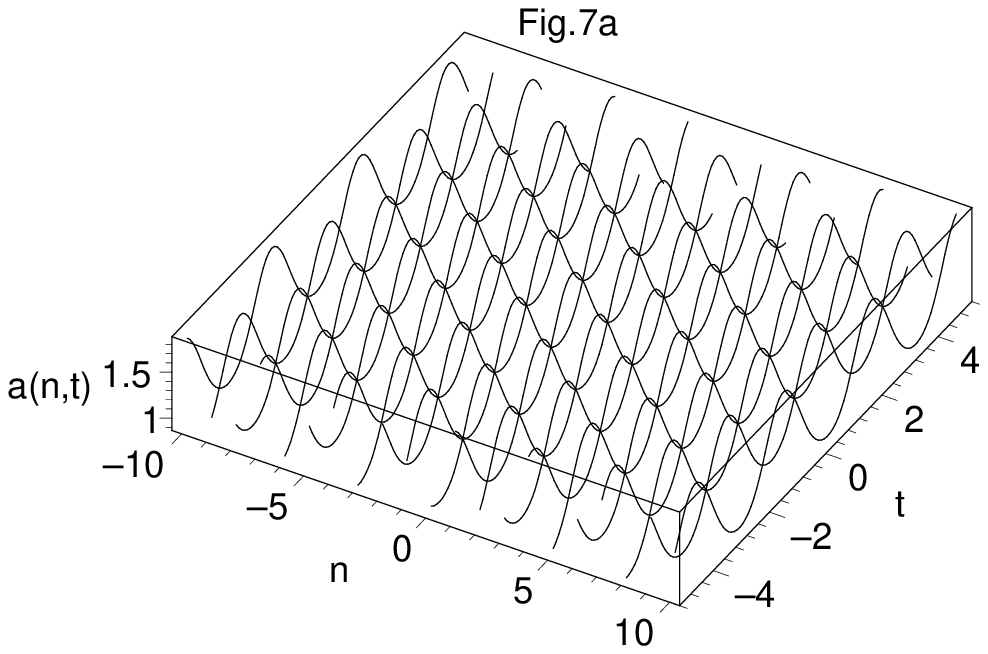}
\epsfxsize=7cm\epsfysize=5cm\epsfbox{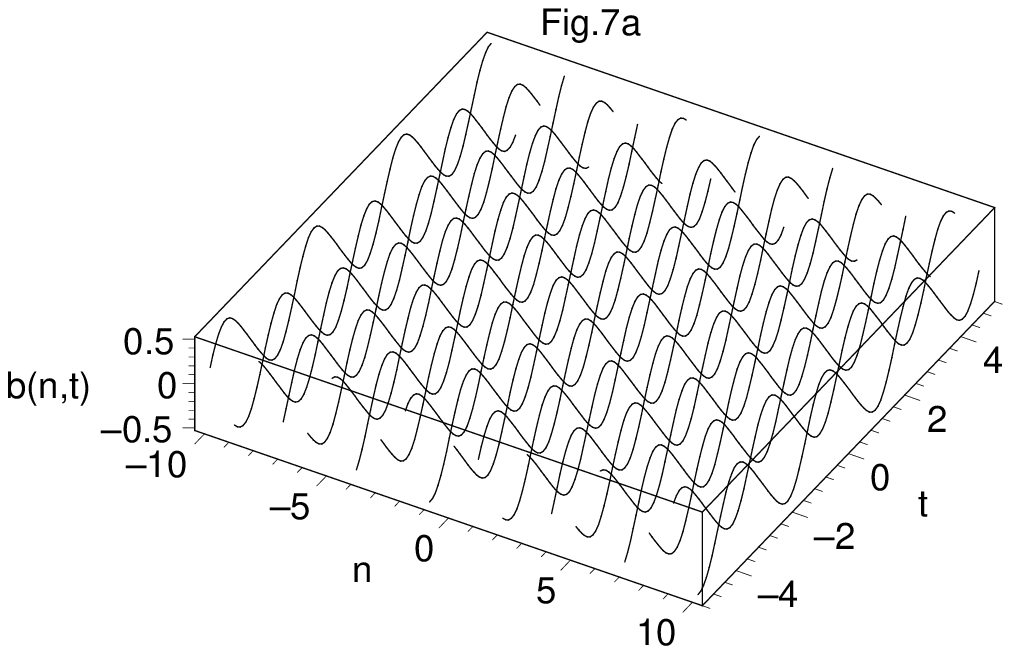}
 \caption{The structure of the positon expressed by (a) \eqref{Sol1} and (b) \eqref{Sol2}
 with \eqref{d01}--\eqref{b21} and the parameter selections \eqref{para3}.}
\end{figure}

To get analytical complexiton solutions of the coupled Volterra
system \eqref{cvolt1}-\eqref{cvolt2}, we may use the following
solution ansatz
\begin{eqnarray}
a(n,t)&=&\frac{a_0+a_1\cosh(\xi_1)\cos(\xi_2)+a_2\sinh(\xi_1)\sin(\xi_2)+a_3[\cosh(2\xi_1)+\cos(2\xi_2)]}
{b_0+b_1\cosh(\xi_1)\cos(\xi_2)+b_2\sinh(\xi_1)\sin(\xi_2)+b_3[\cosh(2\xi_1)+\cos(2\xi_2)]},\label{comp1}\\
b(n,t)&=&\frac{d_0+d_1\cosh(\xi_1)\cos(\xi_2)+d_2\sinh(\xi_1)\sin(\xi_2)+d_3[\cosh(2\xi_2)+\cos(2\xi_1)]}
{b_0+b_1\cosh(\xi_1)\cos(\xi_2)+b_2\sinh(\xi_1)\sin(\xi_2)+b_3[\cosh(2\xi_1)+\cos(2\xi_2)]},\label{comp2}
\end{eqnarray}
where
\begin{eqnarray}
\xi_i=k_in+c_it+\xi_{0i},\ i=1,\ 2. \label{xi}
\end{eqnarray}
 After finishing tedious calculations, we find
that \eqref{cvolt1}-\eqref{cvolt2} really possesses analytical
complexiton solutions (an analytical complexiton is just a usual
breather) if $\alpha<0$ while the constants $a_j,\ b_j$ and $c_j$
for $j=0,\ 1, \ 2$ and $3$ should be determined by
\begin{eqnarray}
a_0&=&c_1[\cos(4c)\sinh(2b)-\sinh(6b)-\cos(2c)\sinh(4b)]\nonumber\\
&&+c_2[\sin(2c) \cosh(4b)- \sin(4c)\cosh(2b)- \sin(6c)],\label{a0}\\
a_1&=&\mp 2\left\{[\sinh(5b)\cos(c)-\cos(5c) \sinh(b)
+\cos(c) \sinh(b)+\sinh(3b)\cos(3c)] c_1\right.\nonumber\\
&&\left.+[\sin(c) \cosh(b)-\sin(c)\cosh(5b)
+\sin(5c)\cosh(b)+\sin(3c)\cosh(3b)] c_2\right\},\\
a_2&=&\mp 2\left\{[\sin(3c)\cosh(3b)-\sin(5c)\cosh(b)
+\sin(c)\cosh(b)+\sin(c)\cosh(5b)] c_1\right.\nonumber\\
&&\left.+[\sinh(5b)\cos(c)-\sinh(3b)\cos(3c) -\cos(5c) \sinh(b)
-\cos(c) \sinh(b)] c_2\right\},\\
a_3&=&-2 c_2\cosh(2b) \sin(2c)-2 c_1 \sinh(2b)\cos(2c),
\end{eqnarray}
\begin{eqnarray}
b_0&=&\frac1{\sqrt{-\alpha}}[\cos(2c)(1+2\cosh(4b))+\cosh(2b)(1-2\cos(4c))
+\cosh(6b)-\cos(6c)],\label{b0}\\
b_1&=&\frac{\mp
4}{\sqrt{-\alpha}}\left\{\cosh(b)\cos(5c)+\cosh(b)\cos(3c)
-\cos(c)\cosh(5b)-\cos(c)\cosh(3b)\right\},\\
b_2&=&\frac{\pm 4}{\sqrt{-\alpha}}\left\{-\sinh(b)
\sin(5c)+\sinh(b) \sin(3c)
+\sin(c) \sinh(5b)-\sin(c) \sinh(3b)\right\},\\
b_3&=&\frac{2}{\sqrt{-\alpha}}[\cosh(4b)-\cos(4c)],
\end{eqnarray}
\begin{eqnarray}
d_0&=&-c_2\cos(4c) \sinh(2b)-c_1 \sin(4c) \cosh(2b)-c_1
\sin(6c)\nonumber\\
&&
+c_2 \sinh(6b)+c_1 \sin(2c)\cosh(4b)+c_2\cos(2c) \sinh(4b),\\
d_1&=&\pm 2\left\{[\sinh(5b)\cos(c)-\cos(5c) \sinh(b)
+\cos(c) \sinh(b)+\sinh(3b)\cos(3c)] c_1\right.\nonumber\\
&&\left.+[\sin(c) \cosh(b)-\sin(c)\cosh(5b)
+\sin(5c)\cosh(b)+\sin(3c)\cosh(3b)] c_2\right\},\\
d_2&=&\pm 2\left\{[\cos(5c) \sinh(b)+\cos(c) \sinh(b)
-\sinh(5b)\cos(c)+\sinh(3b)\cos(3c)] c_1\right.\nonumber\\
&&\left.+[\sin(3c) \cosh(3b)-\sin(5c)\cosh(b) +\sin(c)\cosh(b)
+\sin(c)\cosh(5b)] c_2\right\},\\
d_3&=&2 c_2\cos(2c) \sinh(2b)-2 c_1 \sin(2c) \cosh(2b),\label{d00}
\end{eqnarray}
with $$b\equiv \frac12 k_1\delta,\ c\equiv \frac12 k_2\delta$$ and
$k_1,\ k_2,\ c_1,\ \xi_{01}$ and $\xi_{02}$ being arbitrary
constants.

\input epsf
\begin{figure}
\epsfxsize=7cm\epsfysize=5cm\epsfbox{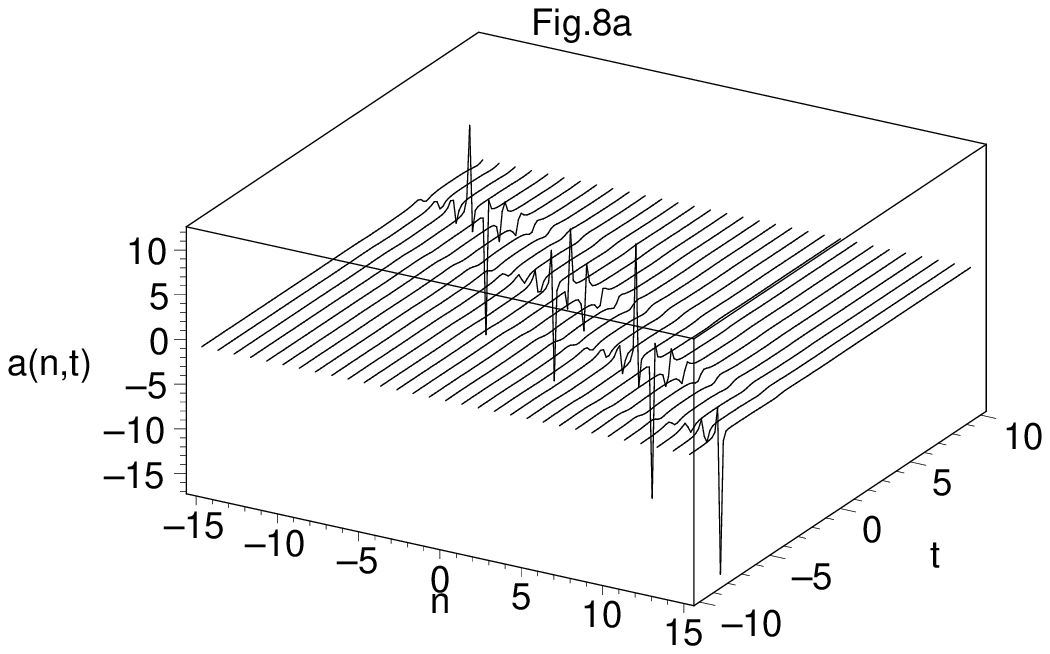}
\epsfxsize=7cm\epsfysize=5cm\epsfbox{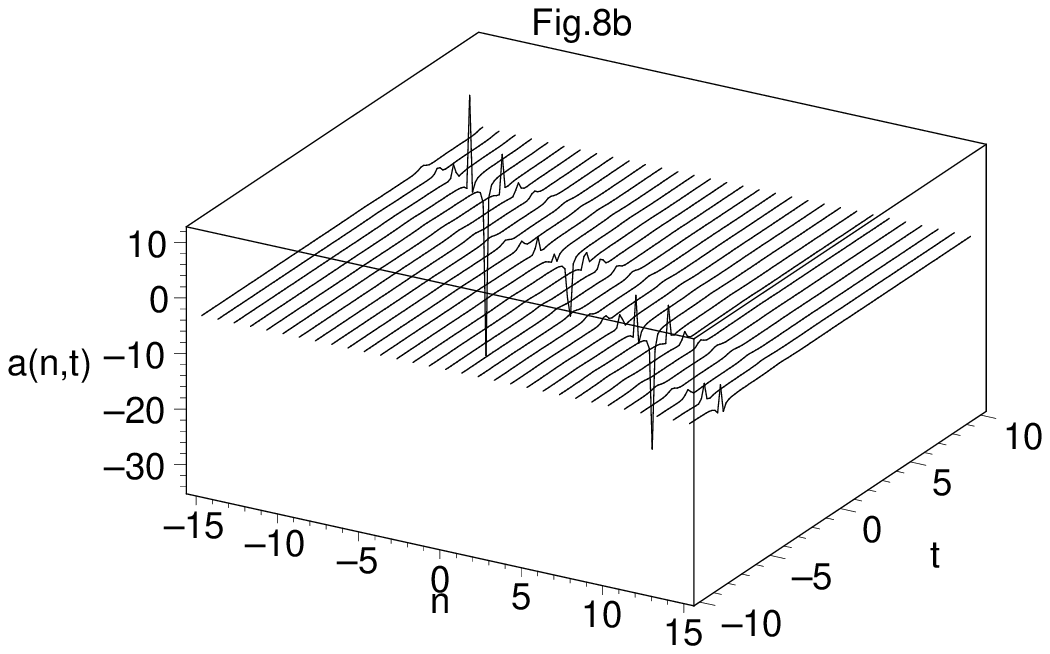}
 \caption{The structure of the complexiton expressed by (a) \eqref{comp1} and (b) \eqref{comp2}
 with \eqref{a0}--\eqref{d00} and the parameter selections \eqref{para4}.}
\end{figure}

Fig. 8 displays the complexiton structure expressed by
\eqref{comp1} and \eqref{comp2}
 with \eqref{a0}--\eqref{d00} and the special parameter selections
\begin{eqnarray}
k_2=\xi_{01}=0,\ k_1=0.1,\ c_1=0.2,\ c_2=0.7, \
\xi_{02}=\frac{\pi}4,\ \delta=1.  \label{para4}
\end{eqnarray}

\section{Summary and discussions}

In summary, the special coupled integrable KdV system
\eqref{ckdv1}--\eqref{ckdv2} is discreterized to an integrable
coupled Volterra system. The Lax integrability of the coupled
Volterra system is proved. By using a simple rational expansion
method of the Jacobi elliptic functions, trigonometric functions
and hyperbolic functions, various exact solutions are found.

For the coupled Vorterra system \eqref{cvolt1}--\eqref{cvolt2}
with $\alpha>0$ there are many types of cnoidal waves described by
different types of Jacobi elliptic functions. On one hand, whence
the modulus of the cnoidal wave tends to 1, the wave tends to a
negaton solution. If a negaton is analytical, then it is called
soliton (or solitary wave for nonintegrable systems). It is found
that only one of these cnoidal waves can be reduced to a single
\emph{analytical} negaton solution when the modulus, $m$, of the
model tends to 1. The soliton solution has a ring or bell shape
for both fields $a(n,t)$ and $b(n,t)$.
 On the other hand, whence the modulus of the
cnoidal wave tends to 0, the wave tends to a positon solution. It
is found that all the nontrivial positons obtained from the
cnoidal waves of section III are singular. This fact is similar to
the KdV and Toda system \cite{Ma}.

In the $\alpha>0$ case, there are two types of single soliton
solutions. In addition to the above mentioned analytical negaton
solution, there is a different type of solitons which has
different shapes for the fields $a(n,t)$ and $b(n,t)$. The new
type of soliton solution is obtained by taking a more complicated
rational expansion of the hyperbolic functions. Because of the
calculation difficulty, we have not yet found any cnoidal wave
extension of this type of negaton solution even utilizing computer
algebras. The field $a(n,t)$ for the second type of soliton
solution also possesses the ring or bell shape while the field
$b(n,t)$ possesses a staggered shape.

For the coupled Vorterra system \eqref{cvolt1}--\eqref{cvolt2}
with $\alpha<0$, though we have not yet found the cnoidal wave
solutions, many other kinds of physically significant solutions,
such as the solitons, analytical positons and analytical
complexitons are found. The structure of the soliton solution in
this case is similar to that of the second type of the solitons
for $\alpha>0$.

It is also interesting that the positon solution can be obtained
by many methods. In this paper, we demonstrate that the analytical
positon solutions of the coupled Vorterra system
\eqref{cvolt1}--\eqref{cvolt2} with $\alpha<0$ can be simply
obtained from the negatons by means of the constant analytical
extensions.

To obtain analytical complexiton solutions of the coupled Vorterra
system \eqref{cvolt1}--\eqref{cvolt2} with $\alpha<0$, a more
complicated rational expansion of both the hyperbolic functions
and trigonometric functions is used.

Finally it is worth to indicated that the real solutions of the
coupled Vorterra system \eqref{cvolt1}--\eqref{cvolt2} with
$\alpha<0$ can also be obtained by means of the analytical
continuous extensions and vice versa. Some analytical continuous
extension examples have been given in section IV. Here, we just
mention a further interesting example. If we apply the constant
transformations
\begin{eqnarray}
k_2\rightarrow \sqrt{-1} k_2,\ c_2\rightarrow \sqrt{-1} c_2,\
\xi_{02}\rightarrow \sqrt{-1} \xi_{02},\ \label{trans}
\end{eqnarray}
to \eqref{comp1}--\eqref{comp2}, then the complexiton solution of
 $\alpha<0$ case becomes a two-soliton solution for $\alpha>0$.

Fig. 9 shows the two-soliton interaction related to
\eqref{comp1}--\eqref{comp2} with the transformation \eqref{trans}
under the parameter (after \eqref{trans}) selections
\begin{eqnarray}
k_2=\xi_{01}=0,\ k_1=0.1,\ c_1=0.1,\ c_2=1, \xi_{02}=\frac{\pi}4,
 \delta=1. \label{para5}
\end{eqnarray}

\input epsf
\begin{figure}
\epsfxsize=7cm\epsfysize=5cm\epsfbox{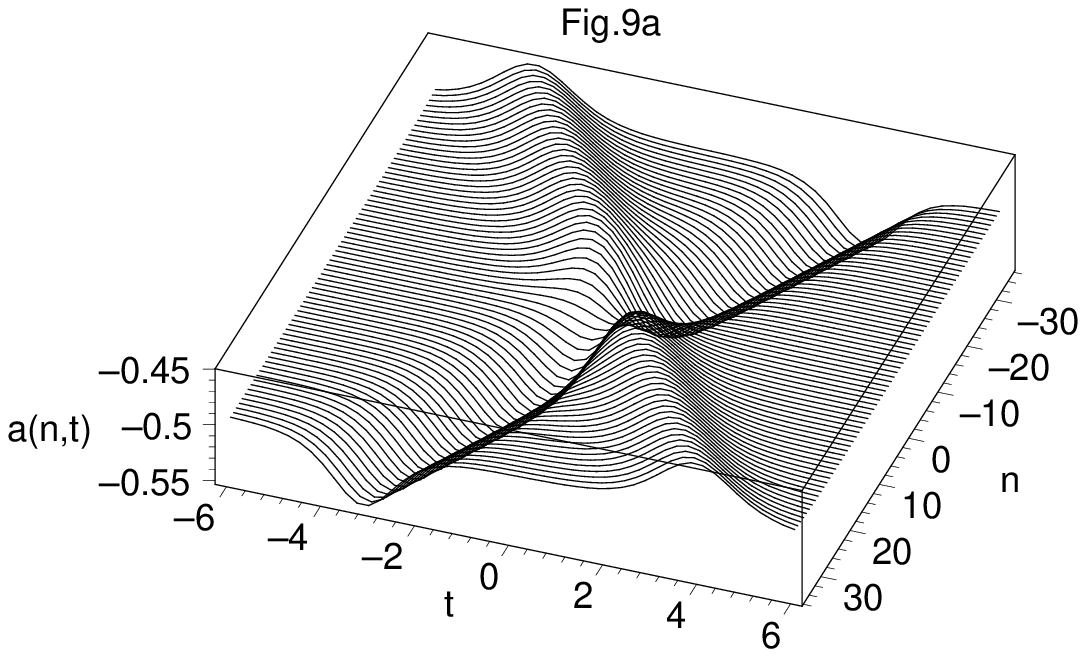}
\epsfxsize=7cm\epsfysize=5cm\epsfbox{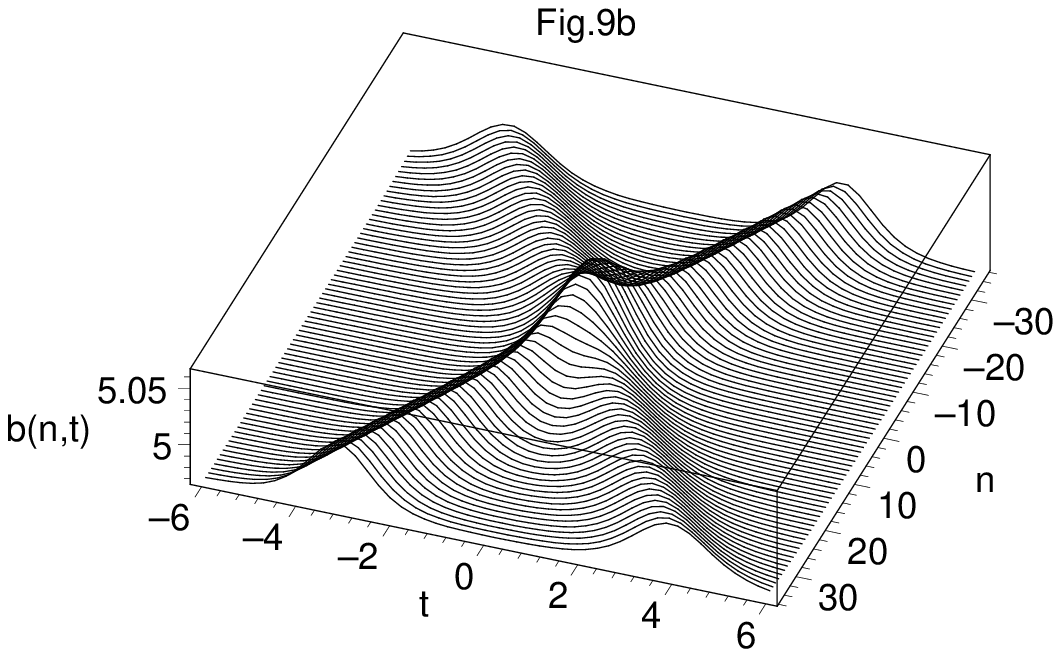}
 \caption{The evolution of the two-soliton interaction solution expressed by (a) \eqref{comp1} and (b) \eqref{comp2}
 with the transformations \eqref{trans} and the parameter selections \eqref{para5}.}
\end{figure}

Though we have obtained many types of exact solutions of the model
via a simple function expansion method, various problems, such as
the general multiple soliton solutions and $\tau$ function
solutions are still open. As a discrete form of the significant
physical model, the more about the model will be studied further.

\begin{acknowledgments}
The authors are in debt to the helpful discussions with Drs. X. Y.
Tang, P. Liu, Y. Gao and X. Y. Jiao. The work was supported by the
National Natural Science Foundations of China (Nos. 10475055,
10735030 and 90503006), the Scientific Research Fund of Zhejiang
Provincial Education Department (No. 20040969) and National Basic
Research Program of China (973 Program 2007CB814800).
\end{acknowledgments}

\end{document}